\documentclass{aa} 

\usepackage{graphicx}
\usepackage{txfonts}
\usepackage{natbib}
\usepackage{url}
\bibpunct{(}{)}{;}{a}{}{,}

\newcommand{\be}{\begin{equation}}
\newcommand{\ee}{\end{equation}}

\def\apjl{ApJL}
\def\apjs{ApJS}

\newcommand{\Mpc}{$h^{-1}$\thinspace Mpc}

\begin{document}  

\title{Tracing high redshift cosmic web with quasar systems 
} 

\author {
Maret~Einasto\inst{1} 
\and Erik Tago\inst{1}
\and Heidi~Lietzen\inst{1,2,3} 
\and Changbom~Park\inst{4}
\and  Pekka~Hein\"am\"aki\inst{5}
\and Enn~Saar\inst{1,6} 
\and Hyunmi~Song\inst{4}
\and Lauri Juhan~Liivam\"agi\inst{1,7} 
\and Jaan~Einasto\inst{1,6,8}  
}

\institute{Tartu Observatory, 61602 T\~oravere, Estonia
\and
Instituto de Astrofísica de Canarias, E-38205 La Laguna, Tenerife, Spain
\and
Universidad de La Laguna, Dept. Astrofísica, E-38206 La Laguna,  
Tenerife, Spain
\and
School of Physics, Korea Institute for Advanced Study, 85 Hoegiro, Dong-Dae-Mun-Gu, Seoul 130-722, Korea
\and 
Tuorla Observatory, University of Turku, V\"ais\"al\"antie 20, Piikki\"o, Finland
\and
Estonian Academy of Sciences,  EE-10130 Tallinn, Estonia
\and
Institute of Physics, Tartu University, T\"ahe 4, 51010 Tartu, Estonia
\and
ICRANet, Piazza della Repubblica 10, 65122 Pescara, Italy
}

\authorrunning{M. Einasto }

\offprints{M. Einasto}

\date{ Received   / Accepted   }

\titlerunning{QSOsystems}

\abstract
{
To understand
the formation, evolution, and present-day properties of the cosmic web
we need to study it at low and high redshifts.   
}
{
We trace the cosmic web at redshifts  $1.0 \leq z \leq 1.8$
using the quasar (QSO) data from the SDSS DR7 QSO catalogue
\citep{2010AJ....139.2360S}.
}
{
We apply a friend-of-friend (FoF) algorithm to 
the quasar and random catalogues to determine systems at a series of linking
lengths, and analyse richness and sizes of these systems. 
}
{At the linking lengths $l \leq 30$~\Mpc\ the number of quasar systems
is larger than the number of systems detected in random
catalogues, and systems themselves have smaller diameters than random
systems. 
The diameters of quasar systems are
comparable to the sizes of poor galaxy superclusters in the local Universe,
the richest quasar systems have four members. 
The mean space density of quasar systems, 
$\approx 10^{-7}\mathrm{(h^{-1}Mpc)}^{-3}$, is
close to the mean space density of local rich superclusters.
At intermediate linking lengths ($40 \leq l \leq 70$~\Mpc)
the richness and length of quasar systems 
are similar to those derived from random catalogues. 
Quasar system diameters are similar to the sizes of rich superclusters 
and supercluster chains in the local Universe. 
The percolating system which penetrate the whole sample volume
appears in quasar sample at smaller linking
length than in random samples ($85$~\Mpc).
At the linking length $70$~\Mpc\ 
the richest systems of quasars have diameters exceeding
$500$~\Mpc, three systems in our catalogue are the same as
described in \citet{2012MNRAS.419..556C, 2013MNRAS.429.2910C}. 
Quasar luminosities in systems are not correlated with the system richness. 
}
{
Quasar system catalogues at our web pages
serve as a database to search for
superclusters of galaxies and to trace the
cosmic web at high redshifts. 
}


\maketitle

\section{Introduction} 
\label{sect:intro} 

Contemporary cosmological paradigm tells us that the structure in the
Universe--the cosmic web formed and evolved  from  tiny density
perturbations in the very early Universe by hierarchical
growth driven by  gravity \citep[see, e.g.,][and 
references therein]{loeb2008, 2009LNP...665..291V}.
Groups and clusters of 
galaxies and their filaments formed by density perturbations of scale up to 
32~\Mpc, superclusters of galaxies, the largest density enhancements in the
local cosmic web, by larger-scale perturbations, up to 100~\Mpc. 
Still larger-scale density perturbations 
modulate the richness of galaxy systems
\citep{2011A&A...534A.128E, 2011A&A...531A.149S}.

To understand  how the  cosmic web formed and  evolved it  is of
paramount importance  to describe and quantify it at  low and  high redshifts.  
Large galaxy
redshift  surveys  like  SDSS  enable  us to  describe  the  cosmic  web  in our
neighbourhood in detail. 
One source of information about  the cosmic structures  at
high redshifts is the distribution of quasars.
Quasars lie at centres of massive galaxies
\citep[][and references therein]{2002AJ....123.2936H,
2009ApJ...703.1663K, 2009MNRAS.397.1862P, 
2010MNRAS.403.2088L, 2013arXiv1302.1366K,
2013IAUS..295...56F, 2013MNRAS.429....2F, 2014arXiv1402.4300F, 2014arXiv1404.1642K},
being
members of galaxy systems, usually small groups and poor clusters
\citep{2000MNRAS.316..267W, 
2002AAS...200.0512S, 2004MNRAS.347.1241S, 2006MNRAS.371..786C,
2009AJ....137.3533H,
2012ApJ...752...39T, 2013ApJ...778...33U}. Nearby quasars are typically
located 
in a relatively low-density large-scale environment near
superclusters of galaxies \citep{2006MNRAS.371..786C,
2009A&A...501..145L, 2011A&A...535A..21L}.

Galaxy luminosities, morphological types, colors, star formation rates,
and other properties
are closely related to their environment at small and
large scales \citep{1974Natur.252..111E, 1980ApJ...236..351D, 1987MNRAS.226..543E,
1992MNRAS.255..382M, 2007ApJ...658..898P, 2009ApJ...691.1828P,
2009MNRAS.392.1467S, 2011A&A...529A..53T,
2012A&A...545A.104L, 2014A&A...562A..87E}.
The studies of quasar environment in the cosmic web helps us to
understand the relation between galaxy and quasar evolution
and to test the evolutionary  schemes of different type of objects. 

Already decades ago several studies described large systems
in quasar distribution 
\citep[][and references therein]{1982MNRAS.199..683W, 1991MNRAS.249..218C, 
1996MNRAS.282..713K, 2002ApJ...578..708W, 2012MNRAS.419..556C, 
2013MNRAS.429.2910C}
which are known as Large Quasar Groups (LQGs). 
\citet{2004MNRAS.355..385M} found significant $200$~\Mpc\ size
over- and underdensities in the density field determined
by quasars from 2dF QSO Redshift survey.  
LQGs may trace  distant galaxy superclusters \citep{1996MNRAS.282..713K}. 
The large-scale distribution of quasar systems
gives us information about the cosmic web at high redshifts
which are not yet covered by large and wide galaxy surveys. 

The goal of this study is to analyse the distribution of quasars at
redshifts $1.0 \leq z \leq 1.8$.
We choose this redshift range as we are interested in the study of
the cosmic web at high redshifts.   This redshift range was also used
by \citet{2012MNRAS.419..556C, 
2013MNRAS.429.2910C}; with this choice we can compare LQGs found 
in these studies, and our systems.
We shall use \citet{2010AJ....139.2360S} catalogue of quasars,
based on the Sloan Digital Sky Survey Data Release 7  (SDSS DR7). 
We analyse clustering properties of quasars with the FoF algorithm, 
compare them with random distributions, and determine
systems of quasars at a series of linking lengths. 
We present catalogues of quasar systems, and analyse their properties and
large-scale distribution. 

In Sect.~\ref{sect:data} we describe the 
data we used and the method in Sect.~\ref{sect:method}.
In Sect.~\ref{sect:results} we give the results. We discuss the 
results in Sect.~\ref{sect:discussion}, 
and summarise the results
 in Sect.~\ref{sect:summ}.
In Appendix~\ref{sect:qdata} we present
data about the richest quasar systems.

At \url{http://www.aai.ee/~maret/QSOsystems.html}
we present the quasar system catalogues. 
There we present also an interactive pdf file showing the distribution of
quasar systems.

We assume  the standard cosmological parameters: the Hubble parameter $H_0=100~ 
h$ km~s$^{-1}$ Mpc$^{-1}$, the matter density $\Omega_{\rm m} = 0.27$, and the 
dark energy density $\Omega_{\Lambda} = 0.73$.

\section{Data} 
\label{sect:data} 

We adopt a catalogue of  quasars with a spectrum taken as a
normal science spectrum (SCIENCEPRIMARY=1) and a photometric
measurement designated PRIMARY in the BEST photometric
database. 
We select from this catalogue a subsample of quasars in the redshift
interval $1.0 \leq z \leq 1.8$, and apply $i$-magnitude limit
$i = 19.1$.
The full quasar catalogue by \citet{2010AJ....139.2360S}
is not homogeneous;  \citet{2005AJ....129.2047V, 2006AJ....131.2766R} 
\citep[see also][]{2012MNRAS.419..556C}
gave a detailed analysis of the completeness of the
sample and suggested to use these selection limits for
statistical studies. 
In order to reduce the edge effects of our analysis, we limit the
data in the area of SDSS sky coordinate limits $-55  \leq \lambda  \leq 55$
degrees and $-33  \leq \eta  \leq 35$ degrees. 
Our final sample contains data of 22381 quasars.

\begin{figure}[ht]
\centering
\resizebox{0.44\textwidth}{!}{\includegraphics[angle=0]{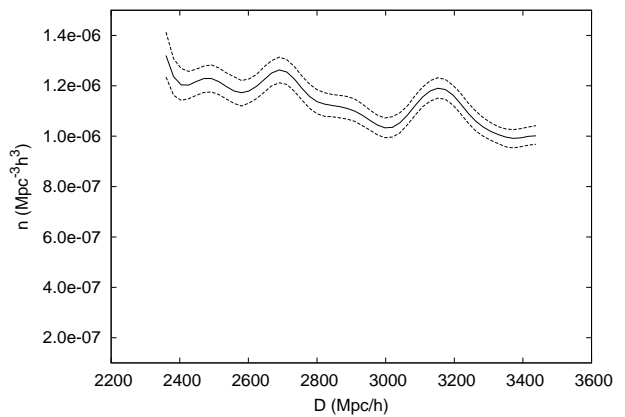}}
\caption{
Quasar number density versus distance. Dotted lines
show the 98\% central confidence limits.
}
\label{fig:distden}
\end{figure}

In Fig.~\ref{fig:distden} we plot the number density of quasars at different
distances. The mean space density of quasars is approximately 
$1.1 \cdot 10^{-6}\mathrm{(h^{-1}Mpc)}^{-3}$. The density decreases slightly
with distance. Below we analyse whether this trend
affects the result of our analysis. 
Figure~\ref{fig:distden} demonstrates several wiggles in the number
density of quasars which correspond to the location  of very rich
quasar systems, as we show below.

The space density of quasars is very low, therefore it is important 
to understand whether their distribution
differs from random distribution.
To compare quasar and
random distributions we generated random samples using the same coordinate  
$\lambda$, $\eta$  and redshift limits as
quasar samples. The number of random points in random samples 
was equal to the number of quasars.

\section{Method}
\label{sect:method} 

We employed the Friends-of-Friends (FoF) clustering analysis 
method introduced in cosmology by \citet{1982Natur.300..407Z} and 
\citet{1982ApJ...257..423H} to determine systems in the quasar catalogue
and random catalogues.
FoF method is commonly used to detect groups and clusters 
in the galaxy distribution \citep[][ and references therein]
{2000ApJS..130..237T, 2004MNRAS.348..866E,
2006AN....327..365T, 2008A&A...479..927T, 2012ApJ...759L...7P,
2012A&A...540A.106T, 2014arXiv1402.1350T}, 
and
superclusters of optical and X-ray  clusters of galaxies \citep{1994MNRAS.269..301E, 
2001AJ....122.2222E, 2013MNRAS.429.3272C}. 
\citet{1996MNRAS.282..713K} applied FoF
method to search for rich quasar systems.

FoF method collects objects into systems if they have at least one 
common neighbour closer than a linking length. At small linking lengths
$l$ only the closest neighbouring objects form systems, most objects
in the sample remain single. As the linking length increases more
and more objects join systems, and the number of systems, their 
richness and size increase. At a certain linking length the largest
system spans over the whole sample volume--a percolation occurs.
We apply FoF algorithm to quasar and random samples with a series
of linking lengths to analyse the properties of systems and their
percolation. 
\citet{1993ApJ...413...48K} showed that the richness and size of the 
richest and 2nd richest system in a sample characterise well the
growth of systems and percolation properties of a sample.
Therefore we analyse the properties of the richest systems in quasar 
and random catalogues at each linking length.

\section{Results}
\label{sect:results} 

\subsection{FoF analysis}
\label{sect:ll} 

The FoF algorithm was applied to observed quasar and random catalogues
at a series
of linking lengths. At each linking length we found the number of systems
in quasar and random samples with at least two members,
calculated multiplicity functions of
systems, and found the length and size of the first and second largest system.
In Tables~\ref{tab:qso203data}, 
\ref{tab:qso304data}, and  \ref{tab:qso7030data} 
we present the data about the richest systems 
of quasars at linking lengths $20$, $30$, and $70$~\Mpc.

\begin{figure}[ht]
\centering
\resizebox{0.45\textwidth}{!}{\includegraphics[angle=0]{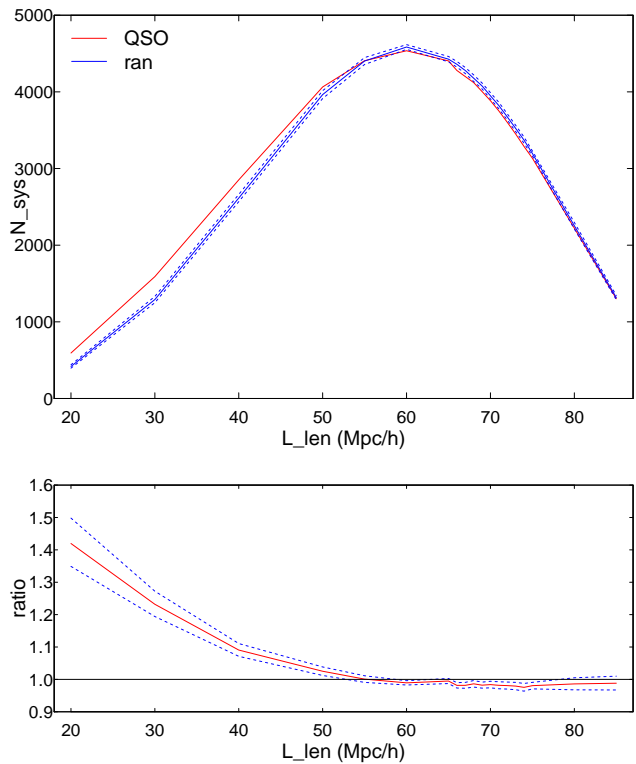}}
\caption{
The number of quasar and random systems (upper panel) and the
ratio of the numbers of quasar and random systems (lower panel)
vs. the linking length. 
Red solid line denote quasar systems, and blue  lines denote random
systems, black line shows ratio 1.
}
\label{fig:nsysll}
\end{figure}

Figure~\ref{fig:nsysll} shows the number of systems from the quasar and random
catalogues (upper panel), and the ratio of the numbers of systems (lower panel). 
At the linking lengths up to approximately $50$~\Mpc\ 
the number of quasar systems is larger than the number of systems in random 
catalogues. At higher values of the
linking lengths the number of systems in quasar
catalogue becomes similar to that in random catalogue.
The number of systems in both quasar and random catalogues
reaches maximum at $60$~\Mpc. At higher values of the linking lengths
systems begin to join into larger systems  and the number of systems 
decreases. 

The richness and diameters (maximum distance between quasar pairs in a system) 
of the largest two systems
in quasar and random catalogues is given 
in Table~\ref{tab:n12nd12}. Table~\ref{tab:n12nd12}
show that at the linking lengths $l \leq 30$~\Mpc\
the richest quasar systems have smaller diameters 
than the richest random systems.  
In a wide linking length interval, $40 \leq l \leq 70$~\Mpc,
the richness and diameters of the richest quasar systems are
comparable to those found in random catalogues.
At the linking length $80$~\Mpc\ quasar systems start to join, 
the richness and diameter of the richest quasar system
increase rapidly.  
The increase of the richness and size of the richest 
random system is much smaller.
At the linking length $l = 85$~\Mpc\ 
about half of quasars join the richest quasar system--
a percolation occurs.  
The richness and diameter of the richest
system in random catalogues also increase, being still poorer
than the richest system in the quasar catalogue.
The second richest system in random catalogues is much richer
than the second richest system in quasar catalogue. This shows that the percolation
of quasar and random systems occurs differently, and the properties of the
richest systems differ.

\begin{table}[ht]
\caption{The number of objects in the two richest systems,  and system diameters
from  quasar and random catalogues
at a series of linking lengths}
\begin{tabular}{rrcrc} 
\hline\hline  
(1)&(2)&(3)&(4)&(5)\\      
\hline 
$l$ & $\mathrm{N1_{QSO}}$ & $\mathrm{N1_{ran}}$ &  $\mathrm{D1_{QSO}}$ & $\mathrm{D1_{ran}}$  \\
\hline
20 &     3 &    3.3$\pm$    0.5 &   31.0 &   33.5 $\pm$   2.0 \\
30 &     4 &    4.9$\pm$    0.8 &   58.8 &   70.0 $\pm$   5.2 \\
40 &     7 &    7.6$\pm$    1.3 &  109.4 &  119.2 $\pm$  17.4 \\
50 &    15 &   12.3$\pm$    2.2 &  178.6 &  209.7 $\pm$  14.7 \\
60 &    23 &   21.3$\pm$    2.0 &  324.9 &  369.8 $\pm$  40.1 \\
70 &    67 &   65.1$\pm$   11.7 &  703.1 &  717.1 $\pm$  86.0 \\
75 &   105 &  146.3$\pm$   39.9 &  865.3 &  992.5 $\pm$ 145.9 \\
80 &   920 &  460.0$\pm$   54.1 & 2066.4 & 1750.2 $\pm$ 171.1 \\  
85 & 11415 & 6464.0$\pm$ 1886.3 & 5060.3 & 4552.7 $\pm$ 645.4 \\  

\hline\hline  
$l$ & $\mathrm{N2_{QSO}}$ & $\mathrm{N2_{ran}}$ &  $\mathrm{D2_{QSO}}$ & $\mathrm{D2_{ran}}$  \\
\hline
20 &   3 &    2.3 $\pm$   0.5 &   29.9 &   30.9 $\pm$  2.6 \\ 
30 &   4 &    3.7 $\pm$   0.5 &   49.4 &   61.4 $\pm$  3.6 \\
40 &   7 &    6.0 $\pm$   0.9 &  106.8 &  104.3 $\pm$  4.9 \\
50 &  14 &   10.6 $\pm$   1.5 &  177.7 &  185.6 $\pm$ 16.3 \\
60 &  20 &   19.9 $\pm$   1.9 &  294.0 &  307.0 $\pm$ 29.3 \\
70 &  64 &   56.3 $\pm$   7.8 &  614.6 &  626.2 $\pm$ 46.9 \\
75 &  97 &  109.1 $\pm$  24.9 &  792.3 &  883.5 $\pm$ 53.8 \\
80 & 421 &  372.4 $\pm$  71.3 & 1832.7 & 1539.5 $\pm$143.0 \\   
85 & 678 & 2652.3 $\pm$ 859.1 & 1789.1 & 2337.6 $\pm$782.2 \\  
\label{tab:n12nd12}                                                          
\end{tabular}\\
\tablefoot{                                                                                 
Columns are as follows:
1: linking length $l$ in \Mpc;
2: the number of quasars in the 1st and 2nd richest quasar system;
3: mean number of random points in the 1st and 2nd richest quasar system;
4: the diameter of the 1st and the 2nd richest quasar system in \Mpc;
5: the diameter of the 1st and the 2nd richest random system in \Mpc.
}
\end{table}

We also analysed the multiplicity functions which show the fractions of 
systems of different richness in quasar
and random samples at a series of linking lengths.
Up to the linking length $l = 70$~\Mpc\ the multiplicity functions for quasar
and random catalogues are very similar. 
At the linking length $80$~\Mpc\ the richness of the richest quasar system
increases rapidly - systems start to join into one huge percolating system,
as we showed above, and multiplicity functions of quasars and random systems become
different. The differences are the largest
at the linking length $85$~\Mpc.

\begin{figure}[ht]
\centering
\resizebox{0.37\textwidth}{!}{\includegraphics[angle=0]{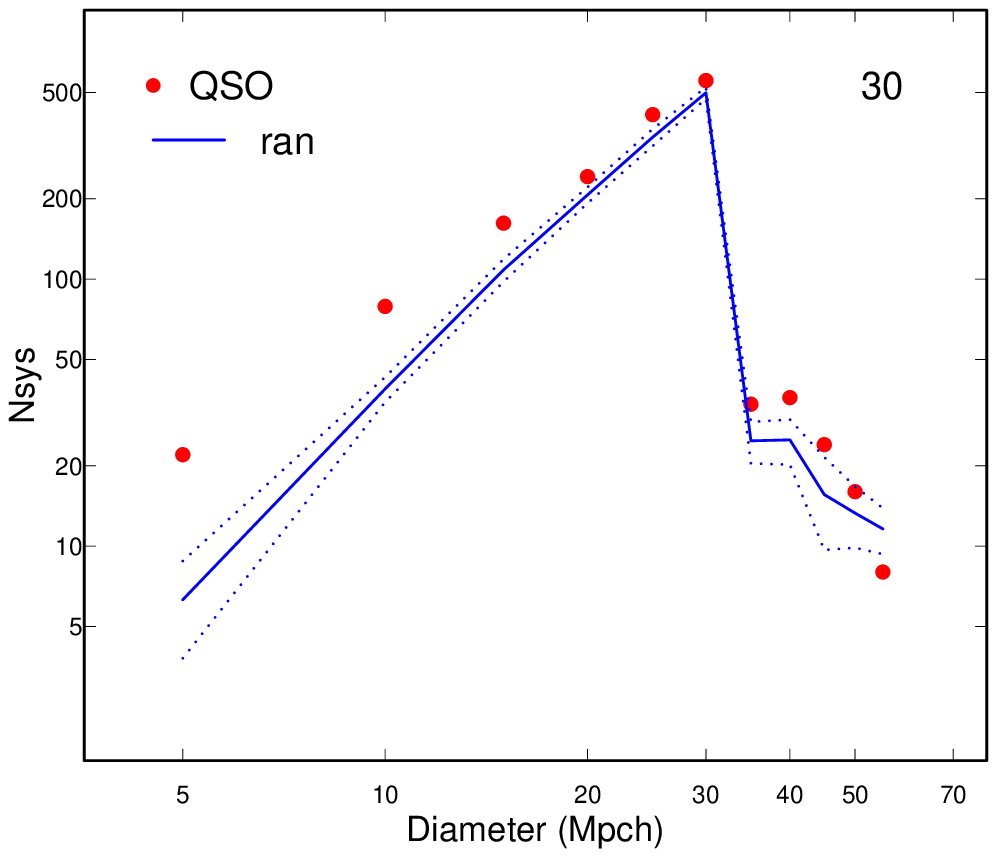}}\\
\resizebox{0.37\textwidth}{!}{\includegraphics[angle=0]{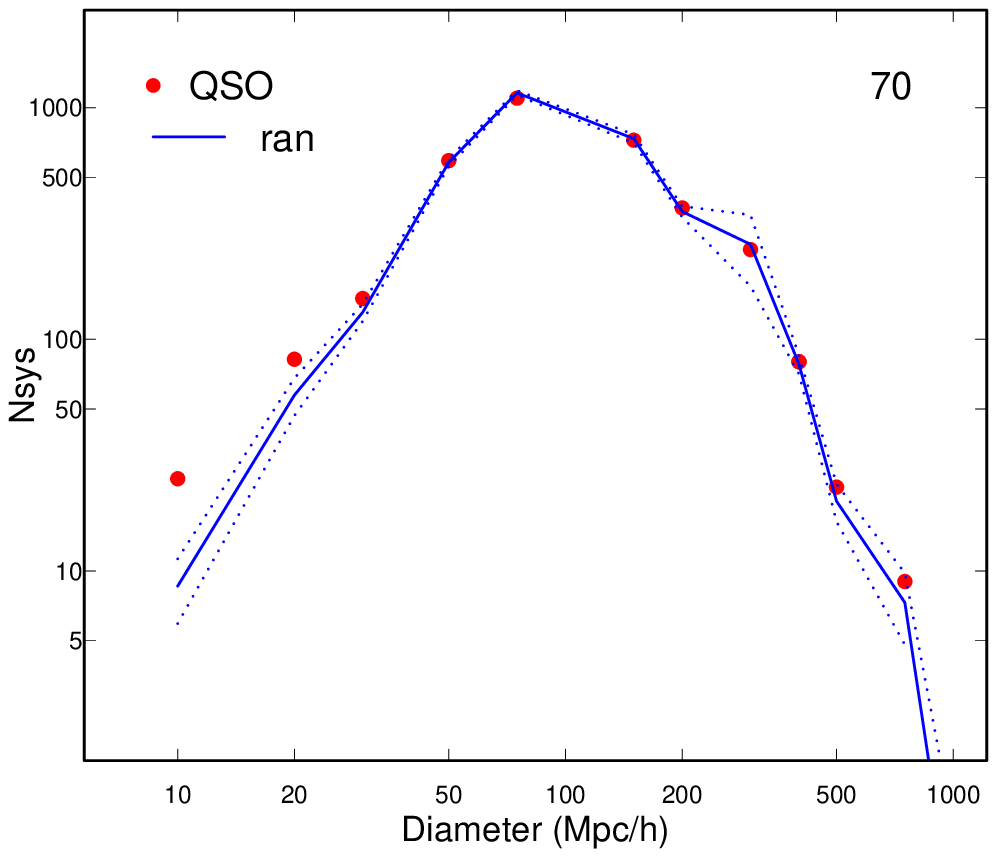}}
\caption{
Number of quasar (red dots) and random (blue lines)
systems of different diameter for linking lengths $30$ and $70$~\Mpc\
(upper and lower panel, correspondingly).
}
\label{fig:diam2070}
\end{figure}

Next we compare the distribution of diameters of quasar and random systems.
We show in Fig.~\ref{fig:diam2070} the numbers of quasar and random
systems of different diameter at the  linking lengths $30$ 
and $70$~\Mpc. At the linking length $20$~\Mpc\ the diameters of the
triple quasar systems are given in Table~\ref{tab:qso203data}.

At the linking lengths $20$ and $30$~\Mpc\ most systems 
in both quasar and random catalogues 
are pairs, 
the richest systems have four (three at $l = 20$~\Mpc) members. The diameters of 
the quasar pairs are as small as  $5$~\Mpc\ in the closest pairs, 
the number of close pairs of quasars
is larger than the number of close random points. The 
lengths of the quasar systems are up to about $30 - 40$~\Mpc,
the highest values of diameters are up to $60$~\Mpc. 
In the whole diameter interval 
the number of quasar systems with a given diameter
is higher than that of random systems (except the highest values of 
diameters; the
number of systems with diameters larger than $50$~\Mpc\
is higher among random systems), the differences are the largest
up to diameters $20$~\Mpc. 

At larger linking lengths ($l \geq 40$~\Mpc; we show this for $70$~\Mpc) 
the number of quasar
systems with diameters up to $20$~\Mpc\ is always larger than the number of
random systems at these diameters. These systems are mostly pairs, but
there are also richer systems among them. Starting from diameters 
$\approx 30$~\Mpc\ the number of systems of different diameter
in quasar and random catalogues becomes similar.

The Kolmogorov-Smirnov (KS) test shows that the differences 
of the distributions of system sizes at the linking lengths $20$ and $30$~\Mpc\
between quasar and random samples are statistically highly significant 
($p_{KS} < 0.003$).  
At higher values of the linking lengths the sizes of quasar systems
with at least three members are statistically similar to those found 
in random samples; the diameters of quasar pairs are at all linking lengths
smaller than diameters of pairs from random catalogues, and these
differences are statistically highly significant ($p_{KS} < 0.05$).

Therefore the FoF analysis reveals that we obtain catalogues of 
real quasar systems in which  the number of systems and 
system diameters  differ from those 
found in random catalogues at  linking lengths $l \leq 30$,
there is relatively more close pairs among quasars than in random
catalogues. This causes also stronger correlation of quasars
than random catalogues at small scales \citep{2007AJ....133.2222S, 
2009MNRAS.394.2050M, 2010MNRAS.407.1078D, 2013ApJ...778...98S}.

\subsection{Properties of quasar systems.}
\label{sect:sysprop}

\begin{figure}[ht]
\centering
\resizebox{0.45\textwidth}{!}{\includegraphics[angle=0]{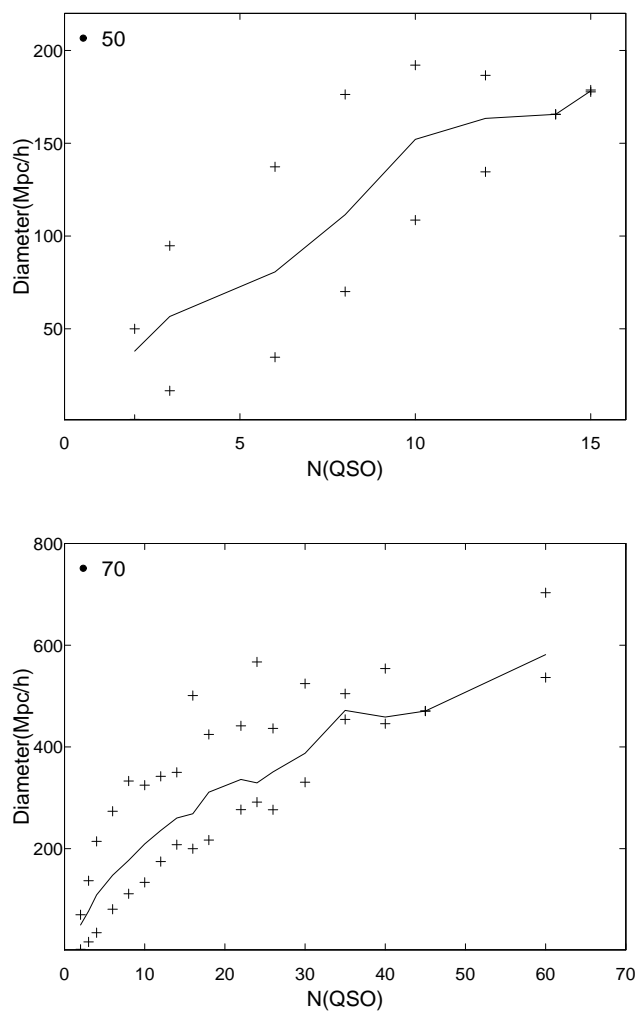}}
\caption{
System richness $N_{QSO}$ vs. their diameter $D_{max}$ for 
quasars for linking lengths $50$ and $70$~\Mpc.
Lines show median values of diameters, crosses denote 
the smallest and the largest diameters.
}
\label{fig:qso70nsysdmax}
\end{figure}

In Fig.~\ref{fig:qso70nsysdmax} we show the median, 
and minimum and maximum values of quasar system 
diameters vs. their
richness at the linking lengths $50$ and $70$~\Mpc.
At the linking lengths $20$ and $30$~\Mpc\ we give the sizes 
of the largest systems 
in Tables~\ref{tab:qso203data} and \ref{tab:qso304data}. 

At the linking lengths $20$ and $30$~\Mpc\
the sizes of quasar systems, up to $30 - 60$~\Mpc, are comparable 
with the sizes of poor and medium rich galaxy superclusters 
in the local universe \citep{1994MNRAS.269..301E, 1997A&AS..123..119E, 
1998A&A...336...35J} with the number of Abell clusters $N_{cl} < 8$.
At the linking lengths $40$ and $50$~\Mpc\ the richest quasar systems
have up to 15 members. The sizes of the richest
quasar systems, $\approx 200$~\Mpc, are comparable to the sizes 
of the richest superclusters and supercluster chains
in the local universe,
typically containing 15--20 rich (Abell) galaxy clusters
\citep{1998A&A...336...35J, 1994MNRAS.269..301E, 
2011MNRAS.411.1716C, 2011A&A...532A..57S,
2012A&A...539A..80L, 2013MNRAS.429.3272C}. 
The mean space density of quasar systems at these linking
lengths is of order of $10^{-7}\mathrm{(h^{-1}Mpc)}^{-3}$, this is
close to the mean space density of rich superclusters
of Abell clusters with at least 10 member clusters
\citep{1997A&AS..123..119E}.
The similarity of the mean space density of quasar systems and 
local rich superclusters was noted already by 
\citet{1996MNRAS.282..713K}.

At the linking length $l = 70$~\Mpc\
rich quasar systems with at least 30 member quasars and size at least
$300$~\Mpc\ (22 systems, Table~\ref{tab:qso7030data})
are comparable in size with those described in
\citet{2012ApJ...759L...7P} and in \citet{2011MNRAS.415..964L} 
as the richest systems in the local universe.
The sizes of the largest quasar systems at $l = 70$~\Mpc, $500 - 700$~\Mpc,
are comparable with the sizes of supercluster complexes, in some
cases penetrating the whole
SDSS survey volume \citep{2011ApJ...736...51E, 2011A&A...532A...5E,
2012A&A...539A..80L}. \citet{2011ApJ...736...51E} 
showed that the richest local supercluster complex, 
the Sloan Great Wall, consists of several rich superclusters.
Individual superclusters in the Sloan Great Wall have different morphology
and galaxy content, therefore they form an assembly of physically different
systems. 

We also analysed luminosities of quasars in systems
of different richness and found that, in average, 
the luminosities of quasars in systems do not depend
on system richness, the luminosities of quasars in systems, and those of
single quasars are similar.

\subsection{Individual  quasar systems.}
\label{sect:qsosys} 

\begin{figure}[ht]
\centering
\resizebox{0.45\textwidth}{!}{\includegraphics[angle=0]{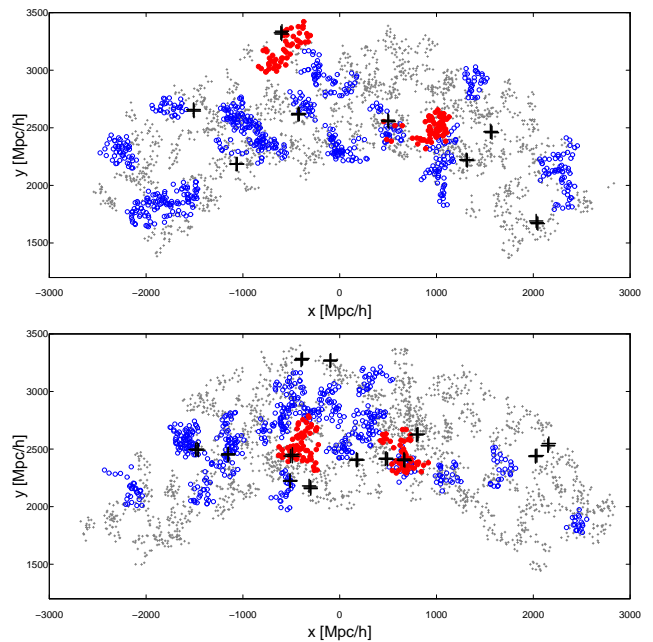}}
\caption{
Distribution of QSO systems at the linking length $70$~\Mpc\ in $x$ and $y$ coordinates
in two slices by $z$ coordinate (upper panel:
$z \leq 0$~\Mpc, lower panel: $z > 0$~\Mpc).
Grey dots denote quasars in systems with $10 \leq N_{QSO} \leq 24$,
blue circles denote quasars in systems with $25 \leq N_{QSO} \leq 49$,
and red filled circles denote quasars in systems with $N_{QSO} \geq 50$.
Black crosses denote quasar triplets at  a linking length $20$~\Mpc.
}
\label{fig:qso70xy}
\end{figure}

We show in Fig.~\ref{fig:qso70xy} the distribution 
of quasars in systems of various richness at linking length $70$~\Mpc\ in 
cartesian coordinates $x$, $y$, and $z$ defined as in
\citet{2007ApJ...658..898P} and in \citet{2012A&A...539A..80L}:
\begin{equation}
\begin{array}{l}
    x = -d \sin\lambda, \nonumber\\[3pt]
    y = d \cos\lambda \cos \eta,\\[3pt]
    z = d \cos\lambda \sin \eta,\nonumber
\end{array}
\label{eq:xyz}
\end{equation}
where $d$ is the comoving distance, and $\lambda$ and $\eta$ are the SDSS 
survey coordinates. 
We plot in the figure also quasars from the richest systems at 
$l = 20$~\Mpc\ (quasar triplets).
The richest  quasar systems at $l = 20$~\Mpc\
are almost all relatively isolated and do not belong to very
rich quasar systems at higher linking lengths.
This can be seen in Table~\ref{tab:qso203data} 
where we give the richness of systems to which they belong at
$l = 70$~\Mpc. Table~\ref{tab:qso203data} shows
that triple quasar systems are located in two relatively
narrow redshift intervals, at $z \approx 1.2$, and at $z \approx 1.7$.

In Fig.~\ref{fig:qso70xy} and Table~\ref{tab:qso7030data} 
four systems at  the linking length $l = 70$~\Mpc\ consist of  more than 
50 quasars, and have diameters larger than $500$~\Mpc,
exceeding the size of the richest local supercluster complex, the Sloan Great Wall.
The richest system at the linking length 
$l = 70$~\Mpc\ at $x \approx 1000$~\Mpc\ and $y \approx 2500$~\Mpc\
is the Huge-LQG described in \citet{2013MNRAS.429.2910C}.
In our catalogue this system has 67 member quasars, and it is
$\approx 700$~\Mpc\ in diameter. 
Below this system, at  $y \approx 2000$~\Mpc\ there are two quasar groups
which coincide with those from \citet{2012MNRAS.419..556C}.

\begin{figure}[ht]
\centering
\resizebox{0.45\textwidth}{!}{\includegraphics[angle=0]{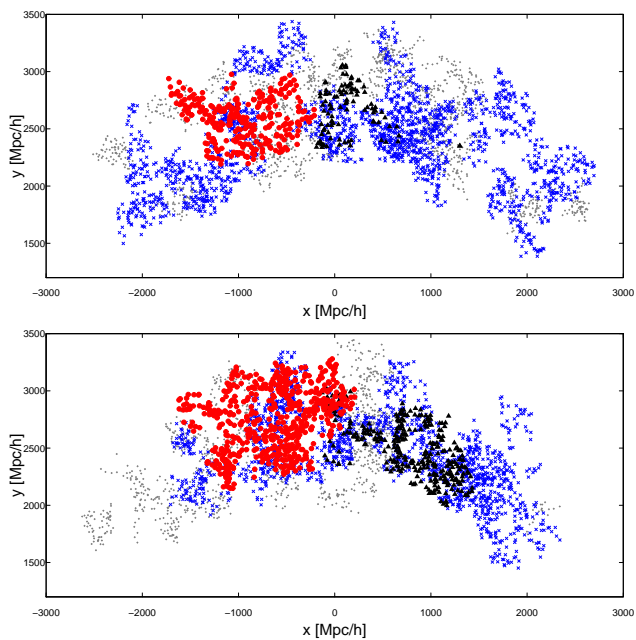}}
\caption{
Distribution of QSO systems at the linking length $80$~\Mpc\ in $x$ and $y$ coordinates
in two slices by $z$ coordinate (upper panel:
$z \leq 0$~\Mpc, lower panel: $z > 0$~\Mpc).
Grey dots denote quasars in systems with $50 \leq N_{QSO} < 100$,
blue circles denote quasars in systems with $100 \leq N_{QSO} \leq 400$,
black triangles denote quasars in the second richest system with $N_{QSO} = 421$
and red filled circles denote quasars in the richest system with $N_{QSO} = 920$.
}
\label{fig:qso80xy}
\end{figure}

Figure~\ref{fig:qso80xy} shows the distribution of quasar systems
at the linking length $80$~\Mpc. The members of the two largest systems are
shown. This figure shows that 
the richest  system form  at $x < 0 $~\Mpc. This is the area where the space density
of quasars at large scales is the highest. 
At this linking length the richest quasar systems found at $l = 70$~\Mpc\
is a member of the second richest system.

\subsection{Large-scale distribution of quasar systems}
\label{sect:lqs}

Visual inspection of Fig.~\ref{fig:qso70xy} shows that in some areas of the figure
there are underdense regions between rich quasar systems
with diameters of about $400$~\Mpc\ (e.q. in the upper panel
between $-1000 < x < 1000$~\Mpc). 
The size of underdense regions in this figure is much larger than
the sizes of typical large voids in the local Universe
\citep[][and references therein]{2011A&A...534A.128E}
but is close to the sizes of the largest voids 
covered by SDSS survey \citep{2011A&A...532A...5E, 2012ApJ...759L...7P}.

\citet{2004MNRAS.355..385M} calculated the density field of quasars from 2dF QSO
redshift survey and showed by visual inspection of the density field plots 
the presence of density fluctuations at scales of about $200$~\Mpc. 
We shall analyse the large scale distribution of quasar systems
in detail in another study.

\section{Discussion and conclusions}
\label{sect:discussion} 

\subsection{Selection effects}
\label{sect:sel}

\begin{figure}[ht]
\centering
\resizebox{0.44\textwidth}{!}{\includegraphics[angle=0]{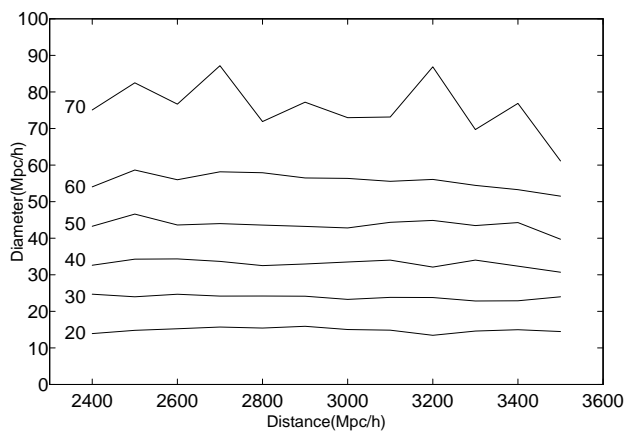}}\\
\caption{
Median diameters of quasar systems at linking lengths $20$--$70$~\Mpc\  
vs. the distance of systems. 
}
\label{fig:diamdist2070}
\end{figure}

We analysed the diameters of systems at different distances 
to see whether the properties of quasar systems found with FoF method are  
affected by distance-dependent selection effects. 
Figure~\ref{fig:diamdist2070} shows the median diameters of quasar systems 
at the linking lengths $20$--$70$~\Mpc\ 
vs. their distances. The median diameters of quasar systems are not 
correlated with distance. We confirmed this also by correlation tests.
We obtained similar results with systems from random catalogues.
Apparent decrease of the median diameters of quasar systems
at the linking length $70$~\Mpc\ at high distances 
is due to an edge effect--close to the sample boundaries FoF cannot find 
very large systems near sample edges. 
This edge effect is much stronger than the influence
of the slight density decrease to the systems properties. 
Due to the edge effect we cannot search for quasar systems dividing
volume under study into narrow slices--this might minimise other
selection effects but makes the edge effects very strong. 
The edge effect affects quasar and random systems in the same way and
may decrease the number of the largest quasar and random
systems at large linking lengths. 
At the linking length $70$~\Mpc\ wiggles in the distribution of the system diameters
are due to the largest quasar systems. These wiggles were
also seen in Fig.~\ref{fig:distden}.

Therefore the slight decrease
of the number density of quasars with distance do not affect significantly
the properties of quasar systems. It is possible that two effects
cancel each other out - the possible increase of quasar systems due to the
decrease of the density, and the decrease of systems due to cosmic
evolution of systems at high redshift \citep{2011A&A...531A.149S}. 
We plan to study quasar and web evolution from simulations
to understand better the evolutionary effects.

With SDSS Visual Tools we checked the images of selected quasars.
We looked at images of all quasars in systems with more than two members at 
the linking lengths  $l = 20$ and $30$~\Mpc, and in the richest systems 
at the linking length $70$~\Mpc, and did not detect contaminated
images.

\subsection{Discussion}
\label{sect:disc}

We found that at the linking lengths $l \geq 40$~\Mpc\ the number of quasar systems,
their sizes and richness are comparable 
with sizes and richness of random systems, and therefore 
their presence do not violate homogeneity of the universe at very large 
scales, as claimed by \citet{2013MNRAS.429.2910C}. 
Similar conclusions were reached by \citet{2013MNRAS.434..398N}.
Transition scale to homogeneity   has been discussed in many studies, see
\citet{1993ApJ...407..443E, 1998MNRAS.298.1212M} and \citet{2013MNRAS.429.2910C} for details and
references.
\citet{2013MNRAS.429.2910C} also mentioned that the richest quasar systems
and therefore the highest large-scale density regions in the Universe are 
located in the area of the Huge-LQG. In our study we found
several quasar systems comparable in richness with the Huge-LQG
in other regions of space, where also the percolating systems appear.
Therefore  the Huge-LQG and other Large Quasar Groups near it
form one of several complexes of rich quasar systems, and not the richest one.

\begin{figure}[ht]
\centering
\resizebox{0.47\textwidth}{!}{\includegraphics[angle=0]{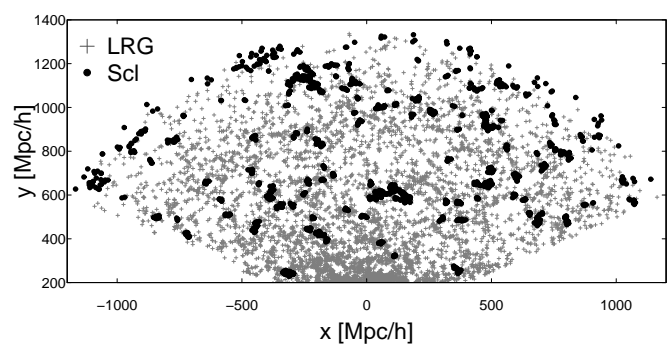}}
\caption{
Distribution of Red Luminous galaxies (light symbols) and their superclusters (dark
symbols) in $x$, $y$ coordinates  in a slice of the z-coordinate 
$-20 \leq z \leq 20$~\Mpc.
}
\label{fig:lrg}
\end{figure}

At the linking length $l = 70$~\Mpc\ the sizes of the largest quasar systems 
are comparable with the sizes of supercluster complexes penetrating the whole
SDSS survey volume \citep{2012A&A...539A..80L}. We illustrate this in
Fig.~\ref{fig:lrg} where we plot the distribution of Luminous Red Galaxies 
and their superclusters from SDSS in $x$, $y$ coordinates \citep{2012A&A...539A..80L}.
An eye can connect the centres of superclusters into huge systems
across the whole volume under study, in analogy with the largest quasar systems.

The mean space density of quasars is approximately 
$1.1 \cdot 10^{-6}\mathrm{(h^{-1}Mpc)}^{-3}$. 
This is much lower than, for example, the mean space density
of nearby galaxy groups in a volume-limited sample
with luminosity limit $M_r <= -21.0$, $2 \cdot 10^{-4}\mathrm{(h^{-1}Mpc)}^{-3}$,
\citep{2014arXiv1402.1350T}.
An analysis of the density enhancements around quasars have shown that, in average,
quasars can be found in overdense environments,
corresponding to small groups or poor (Abell class 0) clusters of galaxies 
with masses $\approx 10^{12}\mathrm{M_{sun}h^{-1}}$
\citep{2000MNRAS.316..267W, 2012ApJ...752...39T, 2013ApJ...778...98S, 
2014MNRAS.441.1802K, 2014arXiv1406.3961M}.
The small space density 
of quasars and mass limit of quasar host haloes tell that
quasars are rare events in small groups, rather than common events in rare,
very rich clusters \citep{2012ApJ...752...39T,
2013ApJ...775L...3T}.
A certain fraction of the group-sized haloes undergo short
\citep[about $10^7$ years, ][]{2014arXiv1406.0778D}
 quasar phase. 
Different physical processes,
which are still not well understood, can trigger
a mass flow onto the central super-massive black
hole (SMBH) \citep{2012MNRAS.420..732P, 2013arXiv1308.5976K}.

\citet{2013MNRAS.436..315F} showed that in simulations the most luminous quasars do not
reside in the most massive dark matter haloes at any redshift. Quasars host haloes
have masses typically about $10^{12}\mathrm{M_{sun}h^{-1}}$, it seems to be quite constant as a
function of 
redshift \citep[see also][and references therein]{2013ApJ...778...98S}. Reason
for this is assumed to be AGN feedback mechanism in haloes with
$M > 10^{12}\mathrm{M_{sun}h^{-1}}$ \citep[][and references therein]{2013MNRAS.436..315F}. 
Simulations indicate that at $z = 6$ the very first quasar haloes 
with mass of $10^{12}\mathrm{M_{sun}h^{-1}}$ are just
forming \citep{2005astro.ph..7197H, 2013MNRAS.436..315F}. The majority of the
descendants of $z = 6$ quasars are located in the haloes which evolve into rich
galaxy clusters in the local Universe 
\citep{2012MNRAS.425.2722A, 2013MNRAS.436..315F}. At lower redshifts
typical quasar host haloes correspond to groups of galaxies.

In the local universe quasars lie in the outskirts of galaxy
superclusters \citep{2009A&A...501..145L, 2011A&A...535A..21L}, and 
in low-density environment
a higher fraction of galaxies hosts AGNs than in high-density 
environment \citep{2004MNRAS.353..713K}.
The clustering of
high-redshift quasars is stronger than the clustering of their low redshift
counterparts \citep{2007AJ....133.2222S}. 

There are only a few galaxy superclusters known
at high redshifts \citep{2005MNRAS.357.1357N, 2007MNRAS.379.1343S, 2009AJ....137.4867L,
2014arXiv1405.2620L}.
The study of the supercluster Cl~1604 at $z = 0.9$ \citep{2008ApJ...677L..89G}
showed that AGNs in this supercluster  avoid the densest environments (the richest
clusters) in the supercluster, being located in 
intermediate density environments in the supercluster, in the outskirts of massive clusters 
or within poorer clusters and groups \citep{2009ApJ...690..295K, 2009ApJ...700..901K}.
The quasar from our sample, at $R.A. = 240.27$ degrees,
$Dec. = 42.74$ degrees, and redshift $z = 1.05$ may be located 
in the outskirts of this supercluster.

The quasar systems found in our study
may mark the location of galaxy superclusters at high redshifts. 
The quasars are typically not 
located in the richest galaxy clusters (in the highest density environments) in
superclusters.
The mean space density of quasar systems at small linking lengths
is of order of $10^{-7}\mathrm{(h^{-1}Mpc)}^{-3}$, this is close to
the mean space density of rich superclusters of Abell clusters
\citep{1997A&AS..123..119E}, in support of this suggestion.

\subsection{Summary and conclusions}
\label{sect:summ} 

We summarise our results as follows: 

\begin{itemize}
\item[1)]
Quasar systems differ from systems found in a random distribution at the
linking lengths $l \leq 30$~\Mpc, their diameters 
(up to $30 - 60$~\Mpc) are 
comparable to the diameters of poor galaxy superclusters in the local Universe.
The richest systems  have four member quasars.
The mean space density of quasar systems, 
$\approx 10^{-7}\mathrm{(h^{-1}Mpc)}^{-3}$, is
close to the mean space density of local rich superclusters.
\item[2)]
In a wide linking length interval, $40 \leq l \leq 70$~\Mpc,
the richness and diameters of the richest quasar systems are
comparable to those found in random catalogues.
The diameters  of richest quasar systems ($\approx 200 - 700$~\Mpc) 
are comparable to the diameters of rich galaxy superclusters and supercluster
chains in the local Universe.
\item[3)]
Quasar systems begin to join into percolating system at the 
linking lengths $l \geq 80$~\Mpc, while in random catalogues the 
percolation occurs at a larger linking length.
\item[4)]
At the linking length $l = 70$~\Mpc\ three systems from our catalogue 
coincide with the Large Quasar Groups from \citet{2012MNRAS.419..556C, 
2013MNRAS.429.2910C}. At this linking length the largest systems 
with diameters exceeding $500$~\Mpc\ are
comparable to the supercluster complexes in the local Universe,
penetrating the whole sample volume.
\end{itemize}

Quasar system catalogues serve as a database to search for distant
superclusters of galaxies and to study the cosmic web at high redshifts.
 
Deeper and wider galaxy surveys are needed to determine quasar's role 
in the evolution of the large-scale structure of the Universe.  
We plan to continue our study of quasars and their systems using
observational data and simulations to understand better how quasars
trace the cosmic web.

\section*{Acknowledgments}

We thank our referee for comments and suggestions which helped to improve the paper.
We are pleased to thank the SDSS Team for the publicly available data
releases.  
Funding for the Sloan Digital Sky Survey (SDSS) and SDSS-II has been
  provided by the Alfred P. Sloan Foundation, the Participating Institutions,
  the National Science Foundation, the U.S.  Department of Energy, the
  National Aeronautics and Space Administration, the Japanese Monbukagakusho,
  and the Max Planck Society, and the Higher Education Funding Council for
  England.  The SDSS Web site is \texttt{http://www.sdss.org/}.
  The SDSS is managed by the Astrophysical Research Consortium (ARC) for the
  Participating Institutions.  The Participating Institutions are the American
  Museum of Natural History, Astrophysical Institute Potsdam, University of
  Basel, University of Cambridge, Case Western Reserve University, The
  University of Chicago, Drexel University, Fermilab, the Institute for
  Advanced Study, the Japan Participation Group, The Johns Hopkins University,
  the Joint Institute for Nuclear Astrophysics, the Kavli Institute for
  Particle Astrophysics and Cosmology, the Korean Scientist Group, the Chinese
  Academy of Sciences (LAMOST), Los Alamos National Laboratory, the
  Max-Planck-Institute for Astronomy (MPIA), the Max-Planck-Institute for
  Astrophysics (MPA), New Mexico State University, Ohio State University,
  University of Pittsburgh, University of Portsmouth, Princeton University,
  the United States Naval Observatory, and the University of Washington.

The present study was supported by the Estonian Ministry for Education and
Science research project SF0060067s08, by ETAG project 
IUT26-2, and by the European Structural Funds
grant for the Centre of Excellence "Dark Matter in (Astro)particle Physics and
Cosmology" TK120. H. Lietzen acknowledges financial support from the Spanish Ministry
of Economy and Competitiveness (MINECO) under the 2011 Severo
Ochoa Program MINECO SEV-2011-0187. This work has also been supported by
ICRAnet through a professorship for Jaan Einasto.
C. Park and H. Song thank Korea Institute for Advanced Study
for providing computing resources (KIAS Center for Advanced
Computation Linux Cluster System).

\bibliographystyle{aa}
\bibliography{qso.bib}

\begin{appendix}

\section{Data on quasar systems}
\label{sect:qdata}

\begin{table*}[ht]
\caption{Data on QSO systems with $N_{QSO} = 3$ at linking length $20$~\Mpc.}
\begin{tabular}{rrrrrrrr} 
\hline\hline  
(1)&(2)&(3)&(4)&(5)& (6)&(7)&(8) \\      
\hline 
 ID& $\mathrm{R.A.}$ & $\mathrm{Dec.}$ &
 $\mathrm{\lambda}$ & $\mathrm{\eta}$  &$\mathrm{z}$ & Diameter & $N_{70}$\\
         &      [deg]  &[deg]   &  [deg] &  [deg]&       &[$h^{-1}$ Mpc] &\\
\hline
 22 & 127.99 & 45.66  & -35.88  &  29.48 &  1.76 &  29.9 &  4 \\
 36 & 132.41 & 38.17  & -38.63  &  19.79 &  1.75 &  22.0 & 10 \\
 53 & 135.41 &  8.50  & -48.85  & -19.51 &  1.21 &  31.0 &  5 \\
124 & 149.80 & 22.13  & -32.27  &  -6.04 &  1.36 &  17.6 &  9 \\
161 & 157.01 &  5.17  & -27.86  & -26.63 &  1.28 &  16.6 &  3 \\
166 & 158.73 & 52.24  & -15.71  &  22.72 &  1.38 &  17.6 &  3 \\
167 & 159.08 & 56.72  & -13.87  &  26.94 &  1.26 &  22.7 &  5 \\
215 & 166.34 & 56.22  & -10.24  &  25.13 &  1.21 &  21.7 &  3 \\
251 & 173.48 & 19.62  & -10.83  & -12.49 &  1.19 &  14.9 &  5 \\
272 & 177.85 & 57.90  &  -3.78  &  25.61 &  1.19 &  19.0 &  9 \\
318 & 187.09 & 36.84  &   1.67  &   4.36 &  1.62 &  22.5 & 17 \\
350 & 193.55 & 37.31  &   6.79  &   5.12 &  1.64 &  20.9 &  3 \\
354 & 194.57 & 19.71  &   9.01  & -12.53 &  1.21 &  18.1 &  3 \\
366 & 196.17 & 24.45  &  10.15  &  -7.63 &  1.72 &  27.8 &  5 \\
379 & 198.68 & 58.00  &   7.20  &  26.23 &  1.04 &  29.2 &  8 \\
415 & 205.47 & 60.14  &  10.02  &  29.23 &  1.30 &  24.2 &  9 \\
430 & 207.98 &  2.79  &  22.95  & -29.46 &  1.22 &  11.0 &  4 \\
435 & 209.40 & 62.39  &  11.03  &  32.03 &  1.19 &  11.1 & 28 \\
456 & 213.77 & 11.32  &  28.16  & -19.62 &  1.55 &  13.8 & 21 \\
472 & 216.68 & 36.76  &  24.88  &   8.77 &  1.23 &  24.9 &  6 \\
544 & 235.02 & 54.57  &  26.37  &  32.93 &  1.65 &  29.9 & 14 \\
\label{tab:qso203data}                                                          
\end{tabular}\\
\tablefoot{                                                                                 
Columns are as follows:
1: ID of the system;
2--3: system right ascension and declination;
4--5: system SDSS coordinates $\lambda$ and $\eta$;
6: mean redshift of the system;
7: system size;
8: the richness of a systems to which a given system belongs at the linking length
$70$~\Mpc.
}
\end{table*}

\begin{table*}[ht]
\caption{Data on QSO systems with $N_{QSO} = 4$ at linking length $30$~\Mpc.}
\begin{tabular}{rrrrrrrr} 
\hline\hline  
(1)&(2)&(3)&(4)&(5)& (6)&(7)& (8) \\      
\hline 
 ID& $\mathrm{R.A.}$ & $\mathrm{Dec.}$ &
 $\mathrm{\lambda}$ & $\mathrm{\eta}$  &$\mathrm{z}$ & Diameter & $N_{70}$\\
         &      [deg]  &[deg]   &  [deg] &  [deg]&       &[$h^{-1}$ Mpc] & \\
\hline
     50  & 128.06 & 45.73 & -35.79 &  29.50 &  1.77  &  34.97 & 4  \\
    122  & 135.29 &  8.64 & -48.94 & -19.27 &  1.21  &  44.55 & 5  \\
    158  & 137.94 &  8.32 & -46.41 & -20.37 &  1.53  &  43.11 & 30 \\
    195  & 140.69 & 14.19 & -42.62 & -13.02 &  1.46  &  49.47 & 28 \\
    496  & 162.75 & 57.62 & -11.69 &  27.09 &  1.10  &  36.37 &  8 \\
    501  & 162.57 & 15.90 & -21.52 & -15.36 &  1.25  &  38.32 & 67 \\
    657  & 173.32 & 19.64 & -10.98 & -12.47 &  1.18  &  37.74 &  5 \\
    721  & 178.13 & 57.89 &  -3.64 &  25.57 &  1.19  &  37.44 &  9 \\
   1135  & 208.12 &  2.89 &  23.09 & -29.34 &  1.22  &  34.70 &  4 \\
   1182  & 212.85 & 58.62 &  14.07 &  29.17 &  1.26  &  58.89 & 16 \\
   1477  & 238.18 & 48.25 &  32.21 &  29.36 &  1.74  &  44.47 &  5 \\
   1490  & 239.04 &  7.08 &  53.44 & -20.55 &  1.10  &  34.74 &  4 \\
\label{tab:qso304data}                                                          
\end{tabular}\\
\tablefoot{                                                                                 
Columns are the same as in Table~\ref{tab:qso203data}.
}
\end{table*}

\begin{table*}[ht]
\caption{Data on QSO systems with $N_{QSO} \geq 30$ at linking length $70$~\Mpc.}
\begin{tabular}{rrrrrrrrr} 
\hline\hline  
(1)&(2)&(3)&(4)&(5)& (6)&(7)&(8)& (9)  \\      
\hline 
 ID&$N_{\mathrm{QSO}}$& $\mathrm{R.A.}$ & $\mathrm{Dec.}$ &
 $\mathrm{\lambda}$ & $\mathrm{\eta}$  &$\mathrm{z}$ & Diameter  &Comments\\
         &     & [deg]  &[deg]   &  [deg] &  [deg]&       &[$h^{-1}$ Mpc] &  \\
\hline
    356  &  30 & 138.06 &   7.86 & -46.34 & -21.09 &  1.55 &   346.8  & \\
   1044  &  33 & 157.76 &  20.72 & -25.33 &  -9.44 &  1.59 &   336.5  & \\
   1093  &  34 & 161.31 &   3.86 & -23.61 & -28.28 &  1.11 &   403.5  & C3 \\
   1156  &  56 & 166.95 &  33.83 & -14.86 &   2.74 &  1.11 &   548.3  & \\
   1159  &  31 & 162.14 &   5.47 & -22.73 & -26.55 &  1.28 &   389.1  & C2 \\
   1195  &  67 & 163.63 &  14.56 & -20.59 & -16.89 &  1.27 &   703.1  & C1 \\
   1593  &  31 & 176.48 &  38.26 &  -6.66 &   6.09 &  1.28 &   330.5  & \\
   1598  &  32 & 177.43 &  32.67 &  -6.38 &   0.41 &  1.18 &   457.3  & \\
   1771  &  33 & 183.60 &  11.03 &  -1.35 & -21.44 &  1.08 &   524.5  & \\
   2002  &  36 & 189.03 &  44.08 &   2.86 &  11.68 &  1.39 &   504.6  & \\
   2011  &  37 & 189.46 &  20.28 &   4.18 & -12.14 &  1.46 &   454.0  & \\
   2190  &  55 & 196.49 &  27.06 &  10.23 &  -4.93 &  1.59 &   614.6  & \\
   2203  &  64 & 196.42 &  39.90 &   8.78 &   8.00 &  1.14 &   536.4  & \\
   2287  &  31 & 199.49 &  54.73 &   8.30 &  23.16 &  1.48 &   381.9  & \\
   2440  &  41 & 205.31 &  50.40 &  12.65 &  19.76 &  1.39 &   554.1  & \\
   2454  &  31 & 202.67 &  21.32 &  16.41 & -10.20 &  1.08 &   375.2  & \\
   2961  &  30 & 217.23 &  10.45 &  31.63 & -20.20 &  1.68 &   387.5  & \\
   3205  &  38 & 227.57 &  41.44 &  30.42 &  17.69 &  1.54 &   471.7  & \\
   3219  &  46 & 226.66 &  16.73 &  39.53 & -10.59 &  1.09 &   470.6  & \\
   3323  &  34 & 228.51 &   8.09 &  42.95 & -21.40 &  1.22 &   399.9  & \\
   3352  &  43 & 230.99 &  47.80 &  28.84 &  25.28 &  1.57 &   458.7  & \\
   3431  &  34 & 234.53 &  10.74 &  48.33 & -16.14 &  1.24 &   390.5  & \\

\label{tab:qso7030data}                                                          
\end{tabular}\\
\tablefoot{                                                                                 
Columns are as follows:
1: ID of the system;
2: the number of quasars in the system, $N_{\mathrm{QSO}}$;
3--4: system right ascension and declination;
5--6: system SDSS coordinates $\lambda$ and $\eta$;
7: mean redshift of the system;
8: system size;
9: comments. C1--C3: large QSO groups, described in: C1--\citet{2013MNRAS.429.2910C}, 
C2, C3--\citet{2012MNRAS.419..556C}. 
}
\end{table*}

\end{appendix}

\end{document}